\documentclass[aps,prd,onecolumn,showpacs,superscriptaddress,nofootinbib,preprintnumbers]{revtex4-2}

\usepackage{style}

\newcommand{\bea}{\begin{equation}\begin{aligned}}
\newcommand{\eea}{\end{aligned}\end{equation}}

\newcommand{\causalacronym}{ECT} 
\newcommand{\rc}{\ensuremath{r_{\rm ect}}}

\begin{document}
\vspace{-0.5cm}
\hfill
\vspace{0.5cm}

\preprint{FERMILAB-PUB-25-0964-T}

\title{\textbf{A Universal CMB $B$-Mode Spectrum from Early Causal Tensor Sources}}

\author{Kylar Greene\,\orcidlink{0000-0002-2711-7191}}
\email{kylar.cosmo@pm.me}
\affiliation{Department of Physics and Astronomy, Seoul National University, 1 Gwanak-ro, Gwanak-gu, Seoul 08826, Korea}

\author{Aurora Ireland\,\orcidlink{0000-0001-5393-0971}} 
\email{anireland@stanford.edu}
\affiliation{Leinweber Institute for Theoretical Physics, Stanford University, Stanford, CA 94305}

\author{Gordan Krnjaic} 
\email{krnjaicg@uchicago.edu}
\affiliation{Fermi National Accelerator Laboratory, Batavia, Illinois 60510}
\affiliation{Kavli Institute for Cosmological Physics, University of Chicago, Chicago, IL 60637}
\affiliation{Department of Astronomy and Astrophysics, University of Chicago, Chicago, IL 60637}


\author{Yuhsin Tsai\,\orcidlink{0000-0001-7847-225X}}
\email{ytsai3@nd.edu}
\affiliation{Department of Physics and Astronomy, University of Notre Dame, South Bend, IN 46556}

\date{\today}

\newcommand{\rlimit}{\ensuremath{0.033}}
\newcommand{\sigmar}{\ensuremath{0.011}}
\newcommand{\aclimit}{\ensuremath{0.0077}}
\newcommand{\sigmaac}{\ensuremath{0.0014}}
\newcommand{\alenslimit}{\ensuremath{1.01^{+0.02}_{-0.02}}}
\newcommand{\sigmaalens}{\ensuremath{0.022}}

\newcommand{\rlimitThreeGSim}{\ensuremath{0.01}}
\newcommand{\sigmarThreeGSim}{\ensuremath{0.0033}}
\newcommand{\aclimitThreeGSim}{\ensuremath{0.00059}}
\newcommand{\sigmaacThreeGSim}{\ensuremath{0.00022}}
\newcommand{\alenslimitThreeGSim}{\ensuremath{-}}
\newcommand{\sigmaalensThreeGSim}{\ensuremath{-}}

\newcommand{\rlimitThreeGDelensSim}{\ensuremath{0.0047}}
\newcommand{\sigmarThreeGDelensSim}{\ensuremath{0.0015}}
\newcommand{\aclimitThreeGDelensSim}{\ensuremath{0.0003}}
\newcommand{\sigmaacThreeGDelensSim}{\ensuremath{0.00012}}
\newcommand{\alenslimitThreeGDelensSim}{\ensuremath{-}}
\newcommand{\sigmaalensThreeGDelensSim}{\ensuremath{-}}

\newcommand{\rlimitThreeGPlusSim}{\ensuremath{0.01}}
\newcommand{\sigmarThreeGPlusSim}{\ensuremath{0.0032}}
\newcommand{\aclimitThreeGPlusSim}{\ensuremath{0.00029}}
\newcommand{\sigmaacThreeGPlusSim}{\ensuremath{0.00011}}
\newcommand{\alenslimitThreeGPlusSim}{\ensuremath{-}}
\newcommand{\sigmaalensThreeGPlusSim}{\ensuremath{-}}

\newcommand{\rlimitThreeGPlusDelensSim}{\ensuremath{0.0011}}
\newcommand{\sigmarThreeGPlusDelensSim}{\ensuremath{3.4\times 10^{-4}}}
\newcommand{\aclimitThreeGPlusDelensSim}{\ensuremath{1.1\times 10^{-4}}}
\newcommand{\sigmaacThreeGPlusDelensSim}{\ensuremath{2.5\times 10^{-5}}}
\newcommand{\alenslimitThreeGPlusDelensSim}{\ensuremath{-}}
\newcommand{\sigmaalensThreeGPlusDelensSim}{\ensuremath{-}}


\begin{abstract}
Many early universe scenarios predict post-inflationary tensor perturbations from causality-limited, sub-horizon sources. While the microphysical details may differ, as long as these sources are bounded in duration and correlation length, their tensor power spectra exhibit a universal scaling behavior at small wavenumber: $\mathcal{P}_h(k) \propto k^3$, corresponding to white noise on super-horizon scales at the time of production. 
If these early causal tensor sources (ECTs) exclusively produce gravitational waves before redshift $z \sim 10^5$, this scaling is realized on all of the scales observed in the cosmic microwave background (CMB), and thus yields a universal multipole distribution for the $B$-mode angular power spectrum. Unlike the scale-invariant distributions of inflationary $B$ modes, ECTs generically predict enhanced power on small scales and suppressed power on large scales, which allows these source classes to be distinguished given measurements over a sufficient range of angular scales. In this paper, we introduce a unified framework for characterizing ECTs and demonstrate how their universal infrared scaling manifests in low-frequency observables, including CMB $B$ modes and stochastic gravitational wave spectral densities. We illustrate this mapping with representative case studies of this universality class involving first-order phase transitions, topological defects, and enhanced scalar perturbations, which source tensor modes at second order in perturbation theory.


\end{abstract}

\maketitle

\section{Introduction}

Primordial tensor perturbations imprint a distinctive parity-odd ($B$-mode) polarization pattern in the cosmic microwave background (CMB), which offers a unique probe of primordial gravitational waves (GWs)~\cite{kamionkowski1997probe}. 
Since CMB experiments are only sensitive to large ($\gtrsim 10$~Mpc) scales, any candidate source for an observable $B$-mode signal must generate tensor modes with wavelengths of order the horizon shortly before recombination. Cosmological inflation famously satisfies this requirement by predicting a nearly scale-invariant spectrum of tensor perturbations with ample support on CMB scales~\cite{Linde:1981mu,Baumann:2009ds,kamionkowski2016quest}. By contrast, post-inflationary  sources are limited by sub-horizon causality and typically assumed to occur well before the CMB era, so they {\it predominantly} generate tensor modes on inaccessibly small scales. Thus, it is often said that a positive $B$-mode signal offers a conclusive verification of inflation~\cite{Weinberg:2003ur}.\footnote{A scale-invariant spectrum of GWs is not unique to inflation, and can also arise in alternatives such as string gas cosmology and matter bounce scenarios~\cite{Brandenberger:2011eq}.}

However, this conclusion relies on the assumption that post-inflationary sources contribute negligibly on CMB scales, which need not be true. On large scales, even sub-horizon causal sources necessarily possess long-wavelength power in Fourier space. For any bounded source with finite spatial correlations, causality implies a universal white-noise power spectrum in the small-wavenumber limit, $\mathcal{P}_h \propto k^3$. Thus, sources with peak frequencies outside the CMB window may nevertheless yield appreciable $B$-mode signals provided there exists sufficient power in this white-noise tail. 
First-order cosmological phase transitions~\cite{Greene:2024},  topological defects like cosmic strings~\cite{Avgoustidis:2012} with a finite lifetime, and scalar-induced tensor modes~\cite{Ireland:2025} are all examples of post-inflationary sources limited to the sub-horizon regime which can yield appreciable $B$-mode signals on observable scales. 
As experiments push to ever-fainter primordial signals, viable contributions from sub-horizon sources can rival the amplitudes of inflationary predictions, thereby complicating the interpretations of future detections. Primordial $B$-mode signals can arise from various new-physics scenarios whose predictions differ markedly from the scale-invariant ``smoking gun" of  inflationary tensor modes.

In this paper, we define a broad universality class of \textit{early causal tensor} sources (\causalacronym{}s). 
All members of this class (1) exhibit finite spatial correlation lengths and (2) are only active for a finite time before $z \approx 7 \times 10^4$, such that they predict only white-noise power spectra on CMB scales.\footnote{The precise cutoff depends on the angular multipoles under consideration. The figure $z\approx 7 \times 10^4$ corresponds to the multipole range $20< \ell < 2301$ adopted in the analysis of our companion paper~\cite{MainPaper}.} While their microphysical details can vary considerably on smaller scales, due to causality, all ECTs predict the same $B$-mode angular distribution and exhibit identical infrared (IR) scaling in their gravitational wave (GW) spectral densities. 

Furthermore, since nearly all early-universe GW sources are also ECTs, this framework offers a unified description of $B$ modes and their corresponding stochastic GW backgrounds. Specifically, we show that $B$-mode measurements also serve as a probe of causal GWs at the low frequencies sampled in CMB measurements ($f \sim 10^{-17}$ Hz), thereby complementing pulsar timing array (PTA) and interferometer searches in the IR. In our companion paper~\cite{MainPaper}, we present the first-ever limits on ECTs using $B$-mode data from multiple CMB data sets~\cite{zebrowski25}. 


This paper is structured as follows: we begin in Sec.~\ref{sec:causal} by reviewing the conditions required for the universal white-noise scaling characteristic of ECTs; in Secs.~\ref{sec:Bmode} and~\ref{sec:SGWB} we show how this universal shape for the tensor power spectrum on CMB-relevant scales manifests in our two observables of interest: the angular spectrum of $B$-mode polarization, $\mathcal{D}_\ell^{\rm BB}$, and the spectral density of stochastic GWs, $\Omega_{\rm GW}$; finally, in Sec.~\ref{sec:casestudies} we examine three concrete ECT case studies: a supercooled first-order cosmological phase transition (\ref{subsec:FOPT}), secondary gravitational waves from enhanced scalar perturbations (\ref{subsec:SIGW}), and cosmic strings (\ref{subsec:strings}). Despite the distinct microphysical origins of tensor perturbations in these three scenarios, they are all examples of ECTs on CMB scales.

\section{Early Causal Tensor Sources}\label{sec:causal}

Gravitational waves are sourced by transverse-traceless perturbations $h^{\lambda}_{ij}$ of the metric, where $i,j$ are spatial indices and $\lambda = (+,\times)$ are the polarization components. While many different processes can produce such perturbations in the post-inflationary universe, they all share a defining feature: they arise from causal physics operating on sub-horizon scales.\footnote{In principle it is possible for there to be a second phase of inflation-like expansion in the post-inflationary universe where the comoving Hubble radius shrinks and GWs can be produced with nearly scale-invariant spectra over a range of wavenumbers. This kind of scenario would evade the ECT classification presented here.} Here we demonstrate that causality completely fixes the spectral shape of super-horizon modes for any finite-duration source with a locally conserved stress-energy tensor and spatial correlations, which vanish beyond a finite physical scale.

\subsection{Intuitive Causality Argument}

 Consider the tensor  two-point correlation function $\xi(r)$ at time $\tau_\star$, defined according to
\bea
    \langle h_\star^\lambda(\vec{x}) h_\star^{\lambda'}(\vec{x}') \rangle = \frac{\delta_{\lambda \lambda'}}{2} \xi(r) \,,
\eea
where $r = |\vec{x} - \vec{x}'|$. Here $\xi(r)$ and the (dimensionful) power spectrum $P_h(k)$ are simply related through a Fourier transform,
\bea\label{eq:Phxirelation}
    P_h(k) =
    \int d^3 r e^{-i \vec k \cdot \vec{r}} \xi (r)
    =
    4\pi \int_0^\infty dr \, r^2\, j_0(kr)\xi(r) \,,
\eea
where $j_0$ is a spherical Bessel function of the first kind. 
For any sub-horizon source active in the post-inflationary universe, causality implies that spatial correlations on scales $r \gtrsim R_\star$ are negligible, such that $\xi(r)$ is effectively\footnote{Causality only requires that $\xi(r)$ fall off \textit{sufficiently fast} on super-horizon scales, and so Eq.~(\ref{eq:xistep}) is only approximate. See Appendix~\ref{app:UETC} for further detail.} zero beyond a characteristic scale $R_\star$, which is necessarily sub-horizon. Thus, we can generically approximate the spatial correlation function as 
\bea\label{eq:xistep}
    \xi(r) \simeq f(r) \Theta(R_\star - r) \,,
\eea
where $f(r)$ depends on the microphysical details of the source. In the small $k \ll R_\star^{-1}$ limit, one may expand the spherical Bessel function $j_0(kr)$ such that Eq.~(\ref{eq:Phxirelation}) becomes
\bea
    P_h(k \ll k_\star) = c_0 + c_2 k^2 + \cdots \,,
\eea
where the $c_i$ are $k$-independent coefficients, the first few of which are
\bea
\label{eq:coeff}
    c_0 = 4 \pi \! \int_0^{R_\star} dr \, r^2 f(r) \,, \quad c_2 = - \frac{4\pi}{3!}  \int_0^{R_\star}  dr \, r^4 f(r) \,.
\eea
Thus, to leading order in $k$, $P_h(k)$ is a constant in the IR, and so the {\it dimensionless} power spectrum becomes
\bea\label{eq:PhmathcalPhrelation}
    \mathcal{P}_h(k) \equiv \frac{k^3}{2\pi^2} P_h(k) \propto k^3,
\eea
which follows directly from the causality requirement that sufficiently-separated spatial points must be be uncorrelated. Note that the amplitude of the power spectrum depends on the details of the source through integrals of $f(r)$ in Eq.~\eqref{eq:coeff}, but the slope is model-independent.

\subsection{A Formal Argument}

To formalize the argument from the previous section, we expand the transverse-traceless tensor perturbation $h_{ij}(\tau,\vec{x})$ in terms of the polarization components of its Fourier modes $h_\lambda(\tau,\vec{k})$ as
\bea\label{eq:hijFourier}
    h_{ij}(\tau,\vec{x}) = \int \frac{d^3 k}{(2\pi)^3} \sum_{\lambda = +,\times} e_{ij}^\lambda(\hat{k}) h_\lambda(\tau, \vec{k}) \, e^{i \vec{k} \cdot \vec{x}} \,,
\eea
where $e_{ij}^{\lambda = (+,\times)}$ form an orthonormal basis of polarization tensors satisfying $e_{ij}^\lambda e^{ij}_{\lambda'} = \delta_{\lambda'}^\lambda$ and $\sum_{\lambda} e_{ij}^\lambda e^{lm}_\lambda = \Lambda_{ij}^{lm}$. Here $\Lambda_{ij}^{lm}$ is the transverse-traceless projector
\bea
    \Lambda_{ij}^{lm}(\hat{k}) = \frac{1}{2} \left(K_i^l K_j^m + K_j^l K_i^m - K_{ij} K^{lm} \right) \,, 
\eea
and we have defined $K_{ij}(\hat{k}) \equiv \delta_{ij} - \hat{k}_i \hat{k}_j$. Each Fourier mode evolves according to the equation of motion
\bea\label{eq:tensorEOM}
    h_\lambda'' + 2 \mathcal{H} h_\lambda' + k^2 h_\lambda = 16 \pi G a^2 \Pi_\lambda \,,
\eea
where $^\prime$ denotes differentiation with respect to conformal time $\tau$, $a$ is the scale factor, and $\mathcal{H} = a'/a$ is the conformal Hubble rate. The source $\Pi_\lambda(\tau,\vec{k})$ is a polarization component of the transverse-traceless projection of the anisotropic stress tensor
\bea
    \Pi_\lambda(\tau,\vec{k}) = e_\lambda^{ij} \Lambda_{ij}^{lm}(\hat{k}) T_{lm}(\tau,\vec{k}) \,,
\eea
whose functional form is model dependent. At this point, we apply our mild physical assumptions that $\Pi_\lambda$ is bounded in frequency and non-vanishing only for the finite time interval $\tau \in [\tau_i, \tau_f]$. The tensor amplitude $h_\lambda(\tau_f, \vec{k}) \equiv h_f^\lambda(\vec{k})$ when the source shuts off can be obtained by solving Eq.~(\ref{eq:tensorEOM}),
\bea\label{eq:hstarlambda}
    h_f^\lambda(\vec{k}) = \frac{16 \pi G}{a_f} \int_{\tau_i}^{\tau_f} d\tau' \, a^3(\tau') G_k(\tau_f,\tau') \Pi_\lambda(\tau',\vec{k}) \,,
\eea
where $a_f \equiv a(\tau_f)$ and 
 $G_k(\tau,\tilde{\tau})$ is the Green's function, which solves the differential equation~\cite{Caprini:2015}
 \be
 \label{eq:Gk}G^{\prime\prime}_k(\tau,\tilde\tau) + \left(k^2 - \frac{ a''}{a} \right)  G_k(\tau,\tilde\tau) = \delta(\tau - \tilde{\tau})~.
 \ee
 During the radiation dominated era, this function has the simple form
\bea\label{eq:Greensfunction}
    G_k(\tau,\tau') = \frac{\sin\left[k(\tau - \tau') \right]}{k} \,.
\eea
The statistical correlations of the tensor perturbation amplitudes at $\tau = \tau_f$ are encoded in the tensor power spectrum $P_h(k)$, defined in the usual manner
\bea\label{eq:Phdef}
    \langle h_f^\lambda(\vec{k}) h_f^{\lambda'}(\vec{k}')^* \rangle = \frac{\delta_{\lambda \lambda'}}{2} P_h(k) (2\pi)^3 \delta^{(3)}(\vec{k} - \vec{k}') \,,
\eea
where $\langle \cdots \rangle$ denotes the ensemble average and the factor of 2 is introduced such that $P_h = \sum_{\lambda} P_h^\lambda$ is the total power spectrum.\footnote{We assume parity conservation so the two polarizations give identical contributions to the total power spectrum.} 
By substituting Eq.~(\ref{eq:hstarlambda}) into Eq.~(\ref{eq:Phdef}) and using the definition of the unequal-time correlator (UETC) $P_\Pi(\tau_1,\tau_2,k)$,
\bea\label{eq:UETCdef}
    \left\langle \Pi_\lambda(\tau_1,\vec{k}) 
    \,\Pi_{\lambda'}(\tau_2, \vec{k}')^* \right\rangle =  \frac{\delta_{\lambda \lambda'}}{2} P_\Pi(\tau_1,\tau_2,k) (2\pi)^3 \delta^{(3)}(\vec{k} - \vec{k}') ,
\eea
one obtains the dimensionful tensor power spectrum
\bea\label{eq:PhUETC}
    P_h(k) =  \bigg( \frac{16 \pi G}{a_f} \bigg)^2 \! \int_{\tau_i}^{\tau_f} \!\! d\tau_1 \int_{\tau_i}^{\tau_f} \!\! d\tau_2 \, a(\tau_1)^3 a(\tau_2)^3 G_k(\tau_f, \tau_1) G_k(\tau_f, \tau_2) P_\Pi(\tau_1, \tau_2, k) \,.
\eea
For a source which is only active for a finite duration $\Delta \tau$, the Green's function from Eq.~(\ref{eq:Greensfunction}) becomes $k$-independent in the super-horizon limit
\bea\label{eq:Gsmallk}
    G_k(\tau,\tau') \simeq (\tau - \tau') + \mathcal{O}\!\left(k^2\right) \, ,
\eea
where $k \ll (\Delta \tau)^{-1}$. 
In the same regime, the UETC of a causal source also becomes $k$-independent\footnote{In Appendix~\ref{app:UETC}, we derive the general conditions under which the UETC becomes $k$-independent in the super-horizon limit. Eq.~(\ref{eq:UETCk0lim}) follows from causality together with the physicality assumptions outlined at the beginning of Sec.~\ref{sec:causal}.}
\bea\label{eq:UETCk0lim}
    P_\Pi(\tau_1, \tau_2, k ) \simeq P_{\Pi}(\tau_1, \tau_2,0) + \mathcal{O}(k^2) \,,
\eea
so, to leading order in the super-horizon limit, the \emph{dimensionful} tensor power spectrum at the time of production satisfies
\bea\label{eq:Ph_const_synth2}
    P_h(k) \propto \int_{\tau_i}^{\tau_f}d\tau_1 \int_{\tau_i}^{\tau_f}d\tau_2\, (\tau_f-\tau_1)\,(\tau_f-\tau_2)\, P_\Pi(\tau_1,\tau_2,0) = \mathrm{const} + \mathcal{O}(k^2) \,.
\eea
Thus, using Eq.~(\ref{eq:PhmathcalPhrelation}), the dimensionless power spectrum scales as
    $\mathcal{P}_h(k) \propto k^3$.
This universal IR behavior is a simple consequence of causality and finite correlation lengths, which dictate that correlations must be white noise (i.e. uncorrelated) on scales which are super-horizon at the time of production.

\subsection{A Universal Ansatz}

\begin{figure}[t!]
\includegraphics[width=\textwidth]{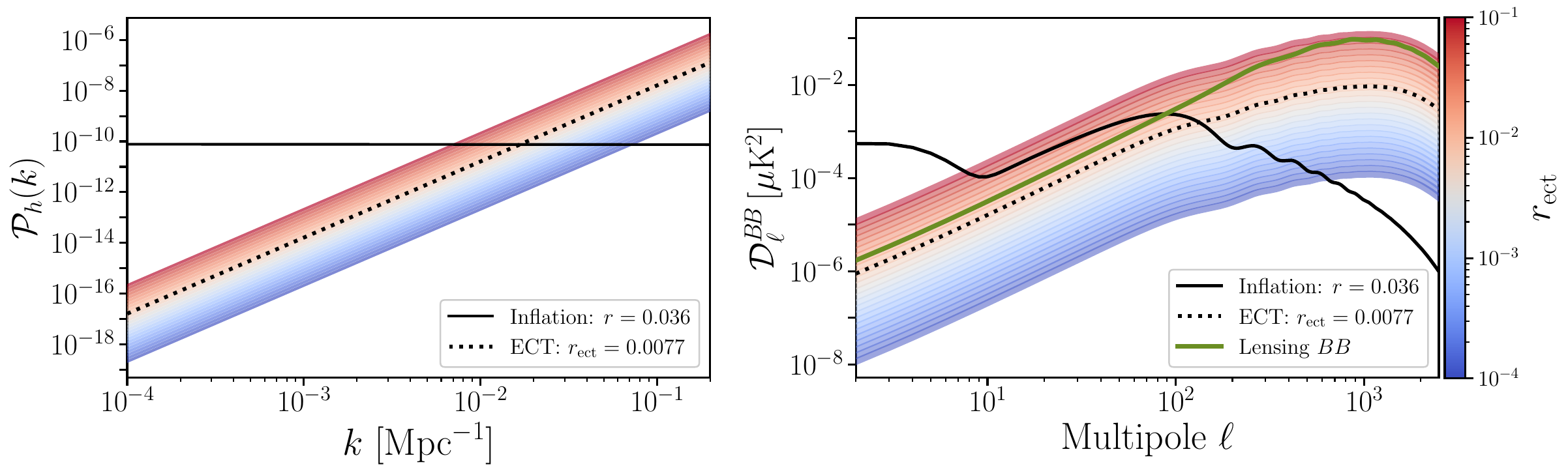}
\caption{Primordial tensor power spectrum $\mathcal{P}_h(k)$ on CMB-relevant scales (left) and resulting CMB $B$-mode spectrum $\mathcal{D}^{BB}_\ell$ (right) for early causal tensor (ECT) sources. The color map encodes the ECT amplitude parameter $r_{\rm ect}$. The solid black curves depict a scale-invariant inflationary reference spectrum with $r=0.036$, chosen to saturate the current limit from BICEP/{\it Keck} \cite{bicepkeck21c}. The lensing $B$-mode prediction for the best-fit Planck 2018 $\Lambda$CDM model is overlaid in green in the right panel. For reference, we include the upper bound of $r_{\rm ect} = 0.0077$ from our companion paper~\cite{MainPaper}, shown in dashed black. 
} 
\label{fig:DlBB}
\end{figure}

Current CMB observables are primarily sensitive to wavenumbers $k_{\rm min} \leq k \leq k_{\rm max}$, where the precise value of $k_{\rm min}$ and $k_{\rm max}$ depends on the angular multipoles under consideration. 
In our companion paper~\cite{MainPaper}, we consider CMB $B$-mode observations spanning $20 < \ell < 2301$, which corresponds to $k_{\rm min}\approx 1\times10^{-3} \, \text{Mpc}^{-1}$ and $k_{\rm max} \approx 0.16 \, \text{Mpc}^{-1}$.\footnote{Importantly, we note that the $k_{\rm min} < k < k_{\rm max}$  range considered here only reflects the sensitivity of current experimental probes and is not a fundamental limit on what can be accessible. As current and future data are analyzed to include higher multipoles, the definition of an ECT must be correspondingly adapted.} Using these reference values, we formalize the definition of an ECT to be any source that exhibits only the causality-limited $k^3$ scaling over the full range of wavenumbers $k_{\rm min} \leq k \leq k_{\rm max}$, which are  relevant for  CMB sensitivity.  
The initial power spectrum of such a source can then be parameterized as
\begin{equation}\label{eq:ansatz}
    \mathcal{P}_h(k) \equiv  \rc{}  A_s \left( \frac{k}{k_{\rm ref}} \right)^3 \,,
\end{equation}
where $A_s = 2.1 \times 10^{-9}$ is the scalar perturbation amplitude at the CMB pivot scale $k_{\rm piv} = 0.05 \, \text{Mpc}^{-1}$~\cite{planck18-6}. We define the parameter \rc{} as the amplitude of the ECT power spectrum at a reference wavenumber $k_{\rm ref}$, measured in units of $A_s$. This normalization is chosen such that limits on \rc{} (derived in~\cite{MainPaper}) can be more intuitively compared with those on inflationary tensor power, conventionally quantified in terms of the tensor-to-scalar ratio $r \equiv A_T/A_s$. For the reference wavenumber, we take $k_{\rm ref} = 0.01 \, \text{Mpc}^{-1}$, since for this value the $B$-mode angular power spectrum from inflation and from an ECT source with $\rc{} = r$ have equal amplitude around the recombination peak ($\ell \sim 100$). In the left panel of Fig.~\ref{fig:DlBB}, we show $\mathcal{P}_h(k)$ for various choices of $r_{\rm ect}$. Later in Sec.~\ref{sec:casestudies}, we demonstrate through concrete case studies that ECT sources asymptote to this universal behavior in the IR.

\section{$B$-Mode Angular Spectrum}\label{sec:Bmode}

The primordial tensor power spectrum enters into the angular spectrum of $B$-mode polarization $\mathcal{D}_\ell^{\rm BB}$ as~\cite{Rubakov:2017}
\bea\label{eq:DlBB}
    \mathcal{D}_\ell^{\rm BB} = 18 \, \ell (\ell+1) \, T_0^2 \int_0^\infty  \frac{dk}{k} \mathcal{P}_h(k) \mathcal{F}_\ell(k)^2 \,,
\eea
where $T_0 = 2.725\, \text{K}$ is the present-day CMB temperature~\cite{Fixsen:2009ug} and we have defined the window function
\begin{equation}\label{eq:Fl}
\begin{split}
    \mathcal{F}_\ell(k) = \int_0^{\tau_0} d\tau \, V(\tau_0,\tau) \mathcal{S}_\ell(\tau,k) \int_0^{\tau} d\tau_1 \, V(\tau,\tau_1) \int_{\tau_1}^\tau d\tau_2 \, \frac{j_2[k(\tau - \tau_2)]}{k^2 (\tau - \tau_2)^2} \frac{\partial \mathcal{T}(\tau_2,k)}{\partial \tau_2} \,,
\end{split}
\end{equation}
where $V(\tau_1,\tau_2)$ is the visibility function, $\mathcal{T}(\tau,k)$ is the tensor transfer function, $j_n(x)$ is a spherical Bessel function of the first kind, and 
\begin{equation}\label{eq:Sfunc}
    \mathcal{S}_\ell(\tau,k)= \frac{\ell + 2}{2\ell + 1} j_{\ell - 1}[k(\tau_0 - \tau)]  -  \frac{\ell-1}{2\ell + 1} j_{\ell + 1}[k(\tau_0 - \tau)] \,,
\end{equation}

The right panel of Fig.~\ref{fig:DlBB} shows the lensing-subtracted $B$-mode spectra for various choices of $r_{\rm ect}$. Because ECTs generate a tensor spectrum $\mathcal{P}_h(k) \propto k^3$ on CMB scales, their $B$-mode signal is suppressed on large angular scales and enhanced on small scales relative to the nearly scale-invariant inflationary prediction. The clearest differences thus occur at low multipoles ($\ell < 10$), where ECTs lack the reionization bump, and at high multipoles, where the ECT signal peaks at $\ell \sim 10^3$ due to the $k$-dependence of the Bessel functions. By contrast, the inflationary spectrum declines rapidly beyond the recombination feature around $\ell \sim 10^2$. Because the ECT and inflationary spectral shapes are distinct, one can in principle distinguish between these sources provided one has $B$-mode measurements across a broad range of angular multipoles. Finally, we note that the angular dependence of the ECT signal closely resembles that of lensing-induced B modes; this degeneracy can be avoided by imposing a prior on the expected lensing power, which can be independently constrained from temperature and E-mode data.

\section{Stochastic GW Background}\label{sec:SGWB}

\begin{figure}
\includegraphics[width=0.65\textwidth]{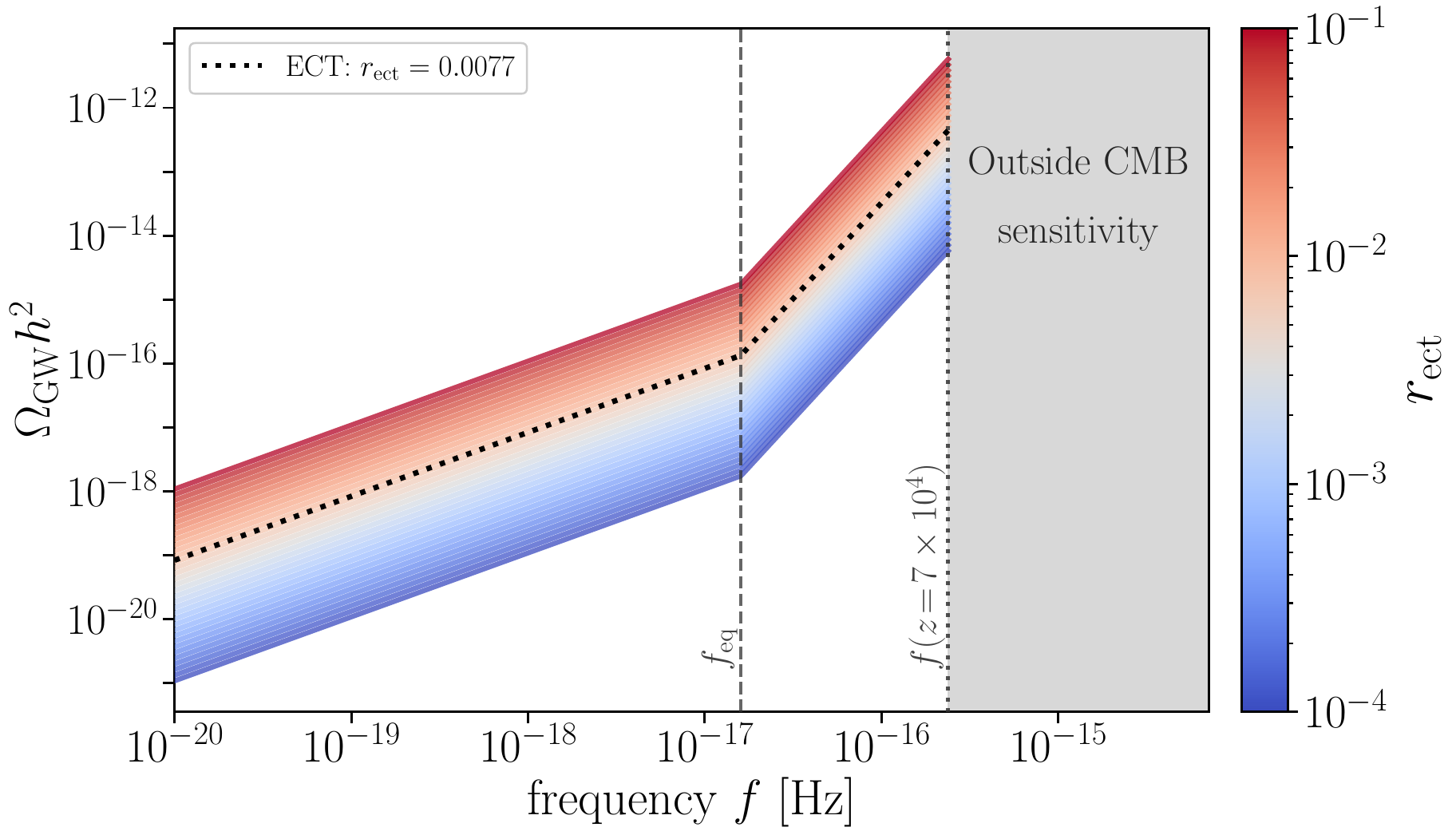}
\caption{Predicted GW spectral density $\Omega_{\rm GW} h^2$ from ECT sources with $r_{\rm ect}\in[10^{-4},10^{-1}]$. The dashed line marks the matter-radiation equality turnover $f_{\rm eq}$. The dotted line bordering the gray region indicates the horizon scale corresponding to $z \simeq 7 \times 10^4$, beyond which the ECT signal lies outside the direct sensitivity of CMB observations. For reference, we include the $\Omega_{\rm GW}h^2$ associated with the upper bound of $r_{\rm ect} = 0.0077$ from our companion paper~\cite{MainPaper}, shown in dashed black.}
\label{fig:OmegaGW}
\end{figure}

Tensor perturbations from ECTs necessarily contribute to the stochastic gravitational wave background (SGWB), and indeed this is the usual context in which ECTs are considered \cite{Niksa:2018ofa,Gouttenoire:2019kij,Chang:2019mza,Hook:2020phx,Gelmini:2021yzu,Auclair:2022ylu,Kitajima:2022lre,Bian:2023dnv,Dimitriou:2023knw,Buckley:2024nen,Caprini:2024ofd,YaserAyazi:2024dxk,Maleknejad:2024ybn,Banik:2024zwj,Balan:2025uke,Baidya:2025ess,Chatrchyan:2025uar,Blasi:2025tmn,Racco:2025ons,Sesana:2025udx,Belyaev:2025zse,Inomata:2024rkt,Bodas:2025wef,Correia:2025qif,Houtz:2025ogg,Dimitriou:2025bvq}. The energy density in GWs is commonly characterized by the dimensionless spectral density
\bea
    \Omega_{\rm GW}(\tau, f) = \frac{1}{\rho_c} \frac{d \rho_{\rm GW}}{d \ln f} \,,
\eea
which represents the GW energy density per logarithmic frequency interval normalized to the critical density $\rho_c = 3 H^2/8 \pi G$. The physical frequency $f$ is related to the comoving wavenumber $k$ through $f = k/2\pi a$. For GWs which have entered the horizon and begun to oscillate, $\rho_{\rm GW}$ is defined as\footnote{The factor 8 in the denominator, as opposed to the usual 32 which appears when writing $\rho_{\rm GW}$ in terms of $h_{ij}$, arises from our convention for the orthonormal basis of polarization tensors -- and hence $h_\lambda$ -- in Eq.~(\ref{eq:hijFourier}).}
\bea
    \rho_{\rm GW}(\tau,\vec{x}) = \frac{1}{8 \pi G a^2} \sum_{\lambda = +,\times} \langle h_\lambda'(\tau,\vec{x})^2 \rangle \,. 
\eea
By working in Fourier space and writing $h' = k h$, 
which is valid at late times after horizon re-entry,
we find
\bea\label{eq:OmegaGWgeneral}
    \Omega_{\rm GW}(\tau, k) = \frac{1}{3} \left( \frac{k}{a H} \right)^2 \mathcal{P}_h(\tau,k) \,.
\eea
Note that ${\cal P}_h(\tau, k)$ is not the initial power spectrum from  Eq.~\eqref{eq:ansatz}, but rather the time-evolved expression obtained by applying the tensor transfer function $\mathcal{T}(\tau,k)$ to $h_\lambda(\tau,\vec{k}) = h_\star^\lambda(\vec{k}) \mathcal{T}(\tau,k)$, which gives
\bea
    \mathcal{P}_h(\tau,k) = \mathcal{P}_h(k) |\mathcal{T}(\tau,k)|^2 \,.
\eea
The transfer function $\mathcal{T}$ solves the sourceless\footnote{More rigorously, free-streaming particles induce a small amount of anisotropic stress, which we neglect for the purposes of this discussion.} version of the perturbation equation of motion from Eq.~\eqref{eq:tensorEOM}, and the full expression in various backgrounds can be found in Refs.~\cite{Turner:1993,Watanabe:2006}. 

Before horizon crossing, Eq.~(\ref{eq:tensorEOM}) is analogous to the equation of motion for an overdamped oscillator, so $\mathcal{T}$ is constant in this regime. After horizon crossing, $\mathcal{T}$ exhibits oscillatory Bessel-like behavior, with the order of the Bessel function determined by the background dynamics during the epoch of re-entry. For fluctuations which were super-horizon at production ($k \ll \mathcal{H}(\tau_f)$) and entered the horizon during an epoch with equation of state $w$, corresponding to a conformal Hubble rate $\mathcal{H} = n/\tau$ where $n = 2/(1+3w)$, the oscillation-averaged scaling is $\mathcal{T} \sim k^{- n}$. Thus, for modes which re-entered during radiation domination $(n = 1)$, $\mathcal{T} \sim 1/k$, and for those which re-entered during matter domination $(n=2)$, $\mathcal{T} \sim 1/k^2$. These properties result in the late-time scaling behavior:
\bea
    \Omega_{\rm GW}(f) \propto \begin{cases} f^3 & \text{radiation domination} \,, \\ f & \text{matter domination} \,. \end{cases}
\eea
Finally, the relevant quantity for the SGWB is the present-day value of the spectral density, which from Eq.~\eqref{eq:OmegaGWgeneral} is
\bea
    \Omega_{\rm GW}(f) = \frac{1}{3} \left( \frac{2 \pi f}{H_0} \right)^2 \mathcal{P}_h(\tau_0, 2 \pi f) \,,
\eea
where $H_0 = 100 \, h$ km/s/Mpc is the present day Hubble rate and we have chosen the normalization $a(\tau_0) = 1$. 

\section{Case Studies}\label{sec:casestudies}

Having established the origin of the universal IR behavior for ECTs and how this manifests in universal shapes for the $B$-mode polarization spectrum and the SGWB spectral density, we now present three well-motivated examples: a supercooled first-order cosmological phase transition, scalar-induced gravitational waves, and cosmic strings with a finite lifetime. Despite their distinct microphysical origins, each spectrum ultimately features the same low-frequency behavior. These case studies both demonstrate that very different scenarios can all fall into the same ECT framework on CMB-observable scales, and clarify where and how the universality assumptions can fail.

\subsection{First-Order Phase Transition}\label{subsec:FOPT}

First-order phase transitions are well-known sources of GWs in the early universe; see Refs.~\cite{Hindmarsh:1994,Croon:2024mde,Caprini:2019} for reviews. 
During such a transition, the system is initially trapped in a false (metastable) vacuum separated from the true vacuum by a potential barrier. As a result, the transition proceeds through the nucleation of true-vacuum bubbles, induced by either quantum tunneling, thermal fluctuations, or stochastic dynamics. As these bubbles grow and collide, they generate anisotropic stress that sources GWs.
While the Standard Model predicts phase transitions associated with electroweak symmetry breaking and hadronic confinement, measured particle properties require both of these transitions to be crossovers, which do not produce GWs~\cite{Laine:2015kra}. 
Thus, observing GWs from a first-order phase transition would be indicative of new physics.

Here we will focus on the case of a strongly supercooled first-order phase transition in a hidden sector containing a scalar field. In such transitions, bubble nucleation completes at a temperature far below the critical temperature at which the minima first become degenerate, $T_n \ll T_c$, where $T$ represents the temperature of the sector undergoing the transition. In this regime, bubble walls typically accelerate to ultra-relativistic velocities $v_w \simeq 1$, and the dominant contribution to the GW signal comes from the gradient energy of the scalar field during the initial bubble collision stage~\cite{Hawking:1982,Kosowsky:1992,Ellis:2019}.

The first-order phase transition provides a particularly transparent realization of ECTs: it is inherently causal, localized in time, and characterized by a correlation length set by the typical bubble size at collision, $R_\star \sim v_w/\beta$,
where $\beta^{-1}$ is the typical timescale from bubble nucleation to collision. When this scale satisfies $k \ll (R_\star/a)^{-1}$ for all CMB-relevant modes, the corresponding tensor power spectrum lies entirely in the causal white-noise regime $\mathcal{P}_h(k) \propto k^3$, and the phase transition falls neatly within the ECT description.

\begin{figure}
\includegraphics[width=0.65\textwidth]{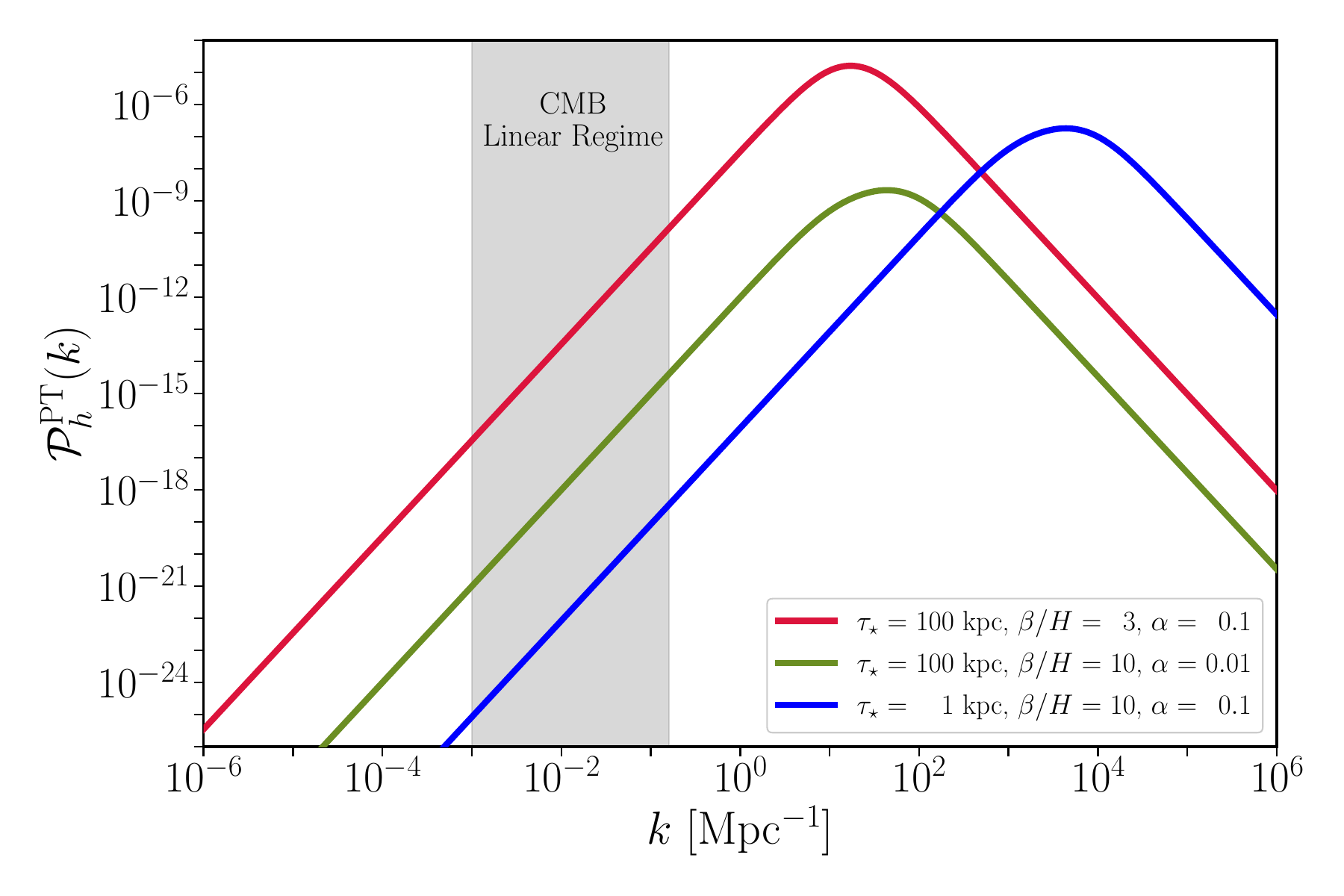}
\caption{Primordial tensor power spectra $\mathcal{P}_h(k)$ sourced by a range of first-order phase transition scenarios in decoupled hidden sectors (colored curves), illustrating the universal causality-limited infrared scaling $\mathcal{P}_h(k)\propto k^3$ at low $k$. These transitions at a characteristic conformal time $\tau_{\star} \sim  \tau_f$, where $\tau = 100$ kpc occurs when the Standard Model photon temperature is 1.1 keV, and $\tau_\star = 1$ kpc corresponds to a photon temperature of 110 keV. Here $\alpha$ is the ratio of false vacuum energy density to the background energy density, following the conventions in \cite{Jinno:2016,Greene:2024}; this parameter only affects the overall normalization of these curves. The shaded gray band indicates the CMB linear regime relevant for observational constraints. 
}
\label{fig:PT_Ph}
\end{figure}

The UETC corresponding to this source $P_\Pi^{\rm PT}(\tau_1,\tau_2,k)$ has been derived in a number of works with varying levels of approximation. The first analytic derivation traces back to Ref.~\cite{Jinno:2016}, which worked in the thin wall and envelope approximations, and also neglected background expansion while the source was still active. Ref.~\cite{Jinno:2017fby} went beyond the envelope approximation, taking into account also bubble propagation and bulk motions after the collisions. Finally, Refs.~\cite{Yamada:2025cfr,Yamada:2025hfs} expanded the calculation to also include the effects of background expansion while the source is active. 
For demonstrative purposes in this section we adopt the thin wall and envelope approximations, though we do account for expansion. Thus, following Ref.~\cite{Yamada:2025cfr}, the single-bubble\footnote{The double-bubble contribution is subdominant, and including it does not affect the conclusions drawn in this section. Nevertheless, we include it in the spectrum of Fig. \ref{fig:PT_Ph}.} contribution to the UETC is
\be
\label{eq:UETCFOPT}
    P_\Pi^{\rm PT}(\tau_1,\tau_2,k) = 4\pi^2 \ell_B^2 \! \int_{|\tau_1 - \tau_2|}^{\tau_1 + \tau_2} \!\!\!\! dr \, r \, \mathsf{P}_2(\tau_1, \tau_2, r) \! \int_{0}^{\tau_{12}} \!\! d\tau_n  \tilde \Gamma(\tau_n) \frac{ \rho_B(\tau_1,\tau_n) \rho_B(\tau_2,\tau_n)}{r_B(\tau_1,\tau_n) r_B(\tau_2,\tau_n)}
    \! \left[  S_0 j_0(kr)  +  S_1 \frac{j_1(kr)}{kr}  +  S_2 \frac{j_2(kr)}{k^2 r^2} \right]\!,~~~~~~ 
\ee
where $\tilde \Gamma$ is the nucleation rate per unit comoving volume, per unit conformal time,  $\tau_{12} = (\tau_1 + \tau_2-r)/2$, $\ell_B$ is the infinitesimal bubble width, and $r_B(\tau,\tau_n) = \tau - \tau_n$ is the comoving bubble radius, where $\tau_n$ is the conformal time of a bubble nucleation event. The function $\rho_B(\tau,\tau_n)$ is the energy density of the bubble wall for points $\vec x$ in the region $r_B < |\vec{x} - \vec{x}_n| < r_B + \ell_B$, where $\vec{x}_n$ is the spatial nucleation point, 
\bea
    \rho_B(\tau,\tau_n) = \frac{1}{a(\tau)^3 r_B(\tau,\tau_n)^2 \ell_B} \int_{\tau_n}^\tau \!\! d\tau' \, a(\tau')^3 r_B(\tau', \tau_n)^2 \kappa(\tau') \rho_0(\tau') \,,
\eea
where $\rho_0$ the vacuum energy and $\kappa$ the efficiency factor, which quantifies how much vacuum energy is converted into energy of the bubble wall. The expressions $S_0$, $S_1$, and $S_2$ are functions of $\tau_{1},\tau_{2},$ and $\tau_n$, and independent of $k$, so their explicit form is not essential for our argument, but can be found in Ref \cite{Yamada:2025cfr}. Finally, following the notation in Ref.~\cite{Yamada:2025cfr}, here   $\mathsf{P} _2(\tau_1,\tau_2,r)$ represents the probability that a pair of spatial points $(\vec{x}_1,\vec{x}_2)$ separated by a distance $r = |\vec x_2 - \vec x_1|$ remain in the false vacuum at conformal times $(\tau_1,\tau_2)$,
\bea
    \mathsf{P}_2(\tau_1,\tau_2,r) = \exp \left( - \int_{V_{12}} d^4 x  \tilde \Gamma(x) \right) \,,
\eea
where the integration is over the region $V_{12} \equiv V_1 \cup V_2$, with $V_i$ is the 4-volume of the past light cone for the point $i$. The dimensionless tensor power spectrum at the end of the phase transition is found by inserting $P_\Pi^{\rm PT}(\tau_1,\tau_2,k)$ from Eq.~\eqref{eq:UETCFOPT} into Eq.~(\ref{eq:PhUETC}) and using ${\cal P}_h = (k^3/2\pi^2) P_h$ to yield
\bea
    \mathcal{P}_h^{\rm PT}(k) \!= \! \frac{k^3}{2\pi^2} \bigg( \frac{16 \pi G}{a_f} \bigg)^2 \! \int_{\tau_i}^{\tau_f} \!\! d\tau_1 \int_{\tau_i}^{\tau_f} \!\! d\tau_2 \, a(\tau_1)^3 a(\tau_2)^3 G_k(\tau_f, \tau_1) G_k(\tau_f, \tau_2) P_\Pi^{\rm PT}(\tau_1,\tau_2,k) \,,
\eea
where $G_k$ is the Green's function from Eq.~\eqref{eq:Gk}.

Figure \ref{fig:PT_Ph} shows the full tensor power spectrum for a range of  benchmark realizations, which all exhibit the expected causality-limited white noise scaling $\mathcal{P}_h^{\rm PT}(k) \propto k^3$ at small wavenumber $k \ll k_f$, where $k_f$ is the comoving wavenumber of the horizon scale at $\tau_f$. The origin of this behavior can also be seen analytically. Noting that the small-argument limits of the spherical Bessel functions appearing in Eq.~(\ref{eq:UETCFOPT}) are 
\be
    j_0(z) \simeq 1 - \frac{z^2}{6} + \cdots \,, \quad \frac{j_1(z)}{z} \simeq \frac{1}{3} - \frac{z^2}{30} + \cdots \,, \quad \frac{ j_2(z)}{z^2} \simeq \frac{1}{15} - \frac{z^2}{210} + \cdots \,,
\ee
so taking the small-$k$ limit, the leading order behavior is 
\bea
    \lim_{k \rightarrow 0} P_\Pi^{\rm PT}(\tau_1,\tau_2,k) = 4\pi^2 \ell_B^2 \int_{|\tau_1 - \tau_2|}^{\tau_1 + \tau_2} \!\!\!\! dr \, r \, \mathsf{P}_2(\tau_1, \tau_2, r) \int_{0}^{\tau_{12}} \!\! d\tau_n  \tilde \Gamma(\tau_n) \frac{ \rho_B(\tau_1,\tau_n) \rho_B(\tau_2,\tau_n)}{r_B(\tau_1,\tau_n) r_B(\tau_2,\tau_n)}  \left( S_0 + \frac{1}{3} S_1 + \frac{1}{15} S_2 \right) \propto k^0,
\eea
which is $k$-independent. Thus, using this same limit of the Green's function from Eq.~\eqref{eq:Gsmallk}, the tensor power spectrum scales as 
\bea
    \lim_{k \rightarrow 0} \mathcal{P}_h^{\rm PT}(k) \!= \! \frac{k^3}{2\pi^2} \bigg( \frac{16 \pi G}{a_f} \bigg)^2 \! \int_{\tau_i}^{\tau_f} \!\! d\tau_1 \int_{\tau_i}^{\tau_f} \!\! d\tau_2 \, a(\tau_1)^3 a(\tau_2)^3 (\tau_f - \tau_1) (\tau_f - \tau_2) P_{\Pi}^{\rm PT}(\tau_1,\tau_2,0) \propto k^3 \,,
\eea
as expected for a causality-limited source in the infrared. From Fig.~\ref{fig:PT_Ph}, we see that for these benchmarks, the $k^3$ white-noise scaling on CMB scales is universal for sufficiently early transitions.

\subsection{Scalar-Induced GWs}\label{subsec:SIGW}

\begin{figure}[t!]
\includegraphics[width=0.65\textwidth]{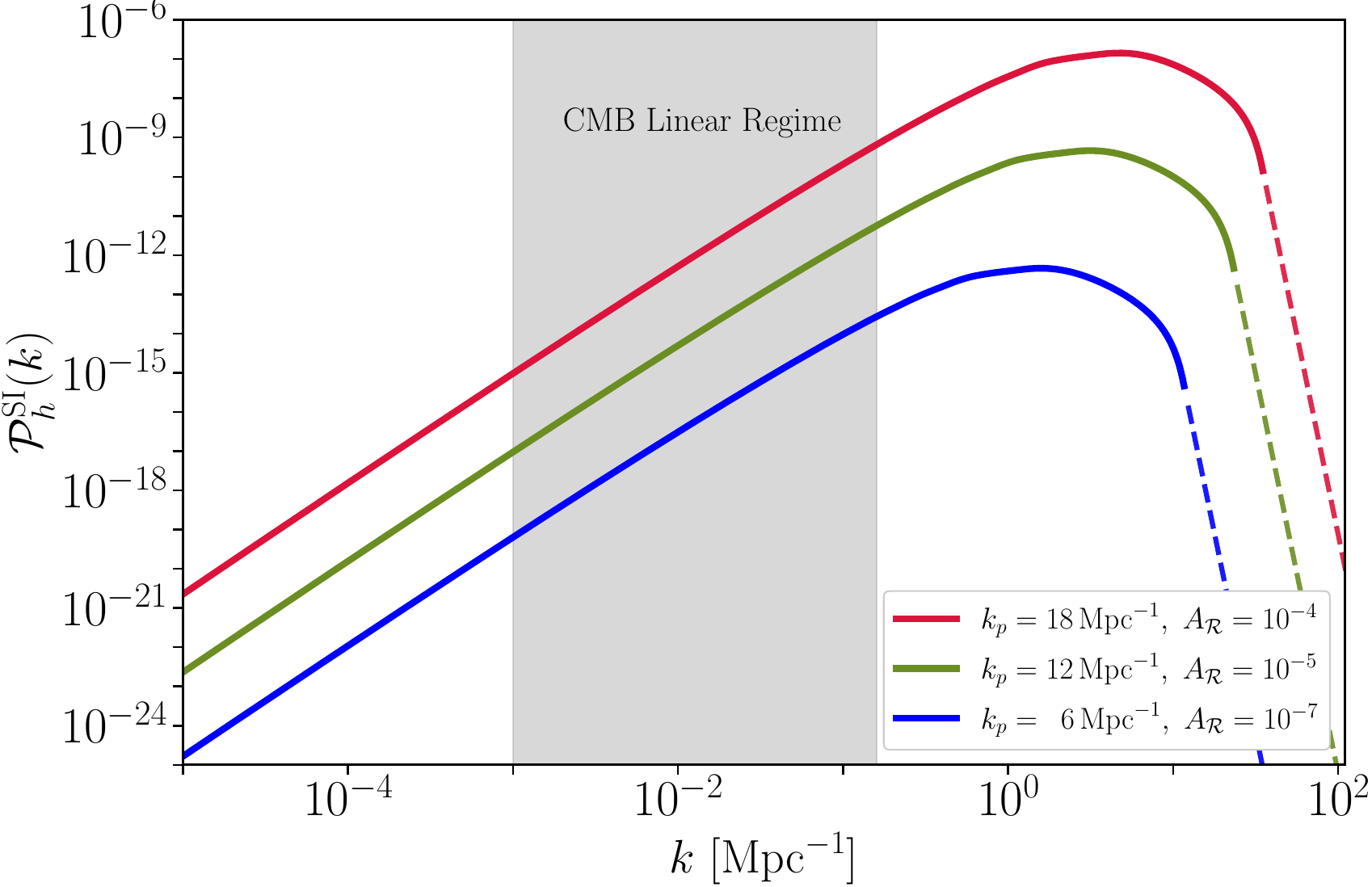}
\caption{The tensor power spectrum $\mathcal{P}^{\rm SI}_h(k)$ sourced at second order by a localized enhancement in the primordial curvature power spectrum, evaluated during radiation domination using the analytic kernel $\mathcal{I}_{\rm RD}$ and a log-box ansatz for $\mathcal{P}_{\mathcal{R}}(k)$. Curves correspond to different choices of the peak wavenumber $k_p$ and amplitude $A_{\mathcal{R}}$. The shaded band indicates the CMB linear sensitivity window. On scales well below the scalar feature ($k \ll k_p$), all cases asymptote to the causality-limited white-noise tail $\mathcal{P}^{\rm SI}_h(k)\propto k^3$. Note that the dashed segments show the expected asymptotic behavior at scaling above $2 k_p$, in accordance with momentum conservation where the scalar source no longer has support to efficiently generate SIGWs.
}
\label{fig:SIGW}
\end{figure}

At second order in cosmological perturbation theory, scalar curvature perturbations $\mathcal{R}$ can source tensor perturbations -- so-called scalar-induced gravitational waves~\cite{Ananda:2006af,Baumann:2007zm,Assadullahi:2009jc}. In standard cosmology, the amplitude of scalar fluctuations is exceedingly small $A_s \sim 10^{-9}$, so a second order effect due to these perturbation is proportional to $A_s^2$ and, therefore, observationally negligible. However, if scalar perturbations are enhanced on small scales, which are outside the nominal range of CMB sensitivity, their second-order effects can be considerable. Indeed, such enhanced scalar perturbations are commonly invoked as a mechanism for primordial black hole formation -- see Ref. \cite{Green:2020jor} for a review.

The scalar-induced source term  in the tensor equation of motion from Eq.~(\ref{eq:tensorEOM})  can be written as \cite{Domenech:2021}
\bea\label{eq:PiSIGW}
    \Pi_\lambda^{\rm SI}(\tau,\vec{k}) = \frac{1}{8 \pi G a^2} \int \frac{d^3 q}{(2\pi)^3} Q_\lambda(\vec{k},\vec{q}) F(\tau,|\vec{k} - \vec{q} \,|,q) \mathcal{R}(\vec{k} - \vec{q}\,) \mathcal{R}(\vec{q} \, ) \,,
\eea
which is indeed quadratic in $\mathcal{R}$. Here, $Q_\lambda(\vec{k},\vec{q}) = e_\lambda^{ij}(\hat{k}) q_i q_j$ is a projection factor, $e_\lambda^{ij}$ is a polarization basis vector defined below Eq.~\eqref{eq:hijFourier}, and we have defined the function
\be
    { F}(\tau, p, q) = \frac{3(1 +w)}{(5 + 3 w)^2} \bigg\{ 2(5\!+\!3w) \phi(p\tau) \phi(q \tau) \!+\! \tau^2 (1\!+\!3w)^2 \phi'(p\tau) \phi'(q\tau) \!+ \!2 \tau (1\!+\!3w) \left[ \phi(p \tau) \phi'(q\tau)\! +\! \phi'(p\tau) \phi(q \tau) \right] \bigg\} ,~~~~~
\ee
where $\phi(q \tau)$ is the scalar-perturbation transfer function and $w$ is the universe's equation of state at the re-entry of mode $q$. The solution to Eq.~\eqref{eq:tensorEOM} for general $\tau$ is given by Eq.~(\ref{eq:hstarlambda}) with the replacements $\tau \rightarrow \tau_f$ and $ \Pi_\lambda \to \Pi_\lambda^{\rm SI}$ from Eq.~\eqref{eq:PiSIGW}. Taking the two-point function of $h_\lambda(\tau,\vec{k})$ and $h_{\lambda'}(\tau,\vec{k}')$ yields
\bea\label{eq:SIGWhh}
    \langle h_\lambda(\tau, \vec{k}) h_{\lambda'}(\tau,\vec{k}') \rangle \!= \!4 \!\int\!\! \frac{d^3 q_1}{(2\pi)^3} \!\int \!\! \frac{d^3 q_2}{(2\pi)^3} Q_\lambda(\vec{k},\vec{q}_1) Q_{\lambda'}(\vec{k}', \vec{q}_2) \mathcal{I}(\tau, & |\vec{k} - \vec{q}_1|, q_1) \mathcal{I}(\tau, |\vec{k}' - \vec{q}_2|, q_2) \\
    & \times \langle \mathcal{R}(\vec{k} - \vec{q}_1) \mathcal{R}(\vec{q}_1) \mathcal{R}(\vec{k}' - \vec{q}_2) \mathcal{R}(\vec{q}_2) \rangle \,,
\eea
where we have defined the combination
\bea
    \mathcal{I}(\tau,p,q) = \frac{1}{a(\tau)} \int_0^\tau \!\! d\tau' \, a(\tau') G_k(\tau,\tau') F(\tau',p,q) \,,
\eea
with the Green's function given by Eq.~\eqref{eq:Greensfunction} for production during radiation domination.\footnote{In the absence of the dissipative effects expected in a realistic cosmological fluid, the source decays only slowly after horizon re-entry, which can lead to a logarithmic running of the $k^3$ tail at small but finite $k$. See Ref.~\cite{Domenech:2025bvr} for a detailed discussion.} 

The 4-point function can be decomposed into connected and disconnected contributions 
\be
\langle \mathcal{R}_{\vec{k}_1} \mathcal{R}_{\vec{k}_2} \mathcal{R}_{\vec{k}_3} \mathcal{R}_{\vec{k}_4} \rangle = \langle \mathcal{R}_{\vec{k}_1} \mathcal{R}_{\vec{k}_2} \mathcal{R}_{\vec{k}_3} \mathcal{R}_{\vec{k}_4} \rangle_c + \langle \mathcal{R}_{\vec{k}_1} \mathcal{R}_{\vec{k}_2} \mathcal{R}_{\vec{k}_3} \mathcal{R}_{\vec{k}_4} \rangle_d.
\ee
The term with the $\langle \cdots \rangle_d$ subscript defines the connected trispectrum $T_\mathcal{R}(\vec{k}_1,\vec{k}_2, \vec{k}_3, \vec{k}_4)$, which vanishes for Gaussian curvature perturbations. For illustration, here we will assume Gaussian statistics, so this term can be neglected.\footnote{See Ref.~\cite{Cai:2018dig,Perna:2024} and references therein for scalar-induced gravitational waves from non-Gaussian curvature perturbations.} The disconnected component can be decomposed according to Wick's theorem, which yields
\be
\langle \mathcal{R}_{\vec{k}_1} \mathcal{R}_{\vec{k}_2} \mathcal{R}_{\vec{k}_3} \mathcal{R}_{\vec{k}_4} \rangle_d = \langle \mathcal{R}_{\vec{k}_1} \mathcal{R}_{\vec{k}_2} \rangle \langle \mathcal{R}_{\vec{k}_3} \mathcal{R}_{\vec{k}_4} \rangle + \langle \mathcal{R}_{\vec{k}_1} \mathcal{R}_{\vec{k}_3} \rangle \langle \mathcal{R}_{\vec{k}_2} \mathcal{R}_{\vec{k}_4} \rangle + \langle \mathcal{R}_{\vec{k}_1} \mathcal{R}_{\vec{k}_4} \rangle \langle \mathcal{R}_{\vec{k}_2} \mathcal{R}_{\vec{k}_3} \rangle,
\ee
so the right-hand side can be simplified using the definition of the scalar power spectrum
\bea
    \langle \mathcal{R}(\vec{k}) \mathcal{R}(\vec{k}')^* \rangle = P_\mathcal{R}(k) (2\pi)^3 \delta^{(3)}(\vec{k} - \vec{k}') \,.
\eea
By performing this replacement in Eq.~\eqref{eq:SIGWhh}, using the delta functions and projection factors to simplify, and comparing with Eq.~(\ref{eq:Phdef}), we finally obtain the scalar-induced tensor power spectrum
\bea
    P_h^{\rm SI}(k) = & 16 \int \frac{d^3 q}{(2\pi)^3} Q_\lambda(\vec{k},\vec{q})^2 \mathcal{I}(\tau,|\vec{k} - \vec{q}\, |, q)^2 P_\mathcal{R}(|\vec{k} - \vec{q} \, |) P_\mathcal{R}(q) \,.
\eea
To evaluate the integral and explore the limiting behavior at small wavenumber, we choose coordinates with $\vec k$ pointing in the $\hat z$ direction where
\be
\vec{k} = k(0,0,1)~~,~~\vec{q} = q(\sin\theta \cos\phi, \sin\theta \sin\phi,\cos\theta),
\ee
and $\sum_\lambda Q_\lambda(\vec{k},\vec{q})^2 = q^4 \sin^4\theta/2$. For concreteness, we assume horizon re-entry during radiation domination, for which $G_k$ is given by Eq.~\eqref{eq:Greensfunction} and the scalar transfer function appearing in ${F}(\tau,p,q)$ is
\be
\phi_{\rm RD}(p\tau) =\frac{3\sqrt{3}}{p\tau} j_1\left(  \frac{p\tau}{\sqrt{3}}\right). 
\ee
Then one can evaluate the time integral $\mathcal{I}(\tau,|\vec{k} - \vec{q}|,q)$ using the expansion $|\vec{k} - \vec{q}|  \simeq q - k\cos\theta$ to find that the leading order contribution at small $k\rightarrow 0$ is $k$-independent, $\mathcal{I}(k \rightarrow 0) = \mathcal{I}_0(q) + \mathcal{O}(k)$. Thus, in terms of the dimensionless curvature power spectrum ${\cal P}_{\cal R} = (k^3/2\pi^2) P_{\cal R}$, we extract the small-$k$ limit
\bea
    \lim_{k \rightarrow 0} \mathcal{P}_h^{\rm SI}(k) = 4 k^3 \!\!\int_0^\infty \! dq \, q^6 \, \mathcal{I}_0(q)^2 \! \int_{-1}^1\!\! dx \, (1-x^2)^2 \frac{\mathcal{P}_\mathcal{R}(|\vec{k}-\vec{q}|)}{|\vec{k} - \vec{q}|^3}  \frac{\mathcal{P}_\mathcal{R}(q)}{q^3}  \,,
\eea
and for any smooth choice of $\mathcal{P}_\mathcal{R}$, we can expand $\mathcal{P}_\mathcal{R}(|\vec{k} - \vec{q}|) \simeq \mathcal{P}_\mathcal{R}(q) - k \cos\theta \mathcal{P}'_\mathcal{R}(q) + \cdots$ to obtain
\bea
    \lim_{k \rightarrow 0} \mathcal{P}_h^{\rm SI}(k) = \frac{64}{15} k^3 \!\int_0^\infty \!dq \, \mathcal{I}_0(q)^2 \mathcal{P}_\mathcal{R}(q)^2 \propto k^3 \,,
\eea
which is indeed the causality-limited white-noise scaling behavior expected from an ECT source. 

In Fig. \ref{fig:SIGW} we show our numerical results for an enhanced scalar power spectrum on small scales. We adopt a log-box ansatz for the dimensionless scalar power spectrum
\be
{\cal P}_{\cal R}(k) =  \begin{cases}
    A_{\cal R} ,   & k \in [ k_p e^{-\Delta/2} ,  k_p e^{+\Delta/2}  ] \\
0,
& \rm otherwise 
\end{cases}~,
\ee
where $A_{\cal R}$ is a constant that sets the scale of the enhanced feature and $k_p$ sets its central position in log $k$ space. For each numerical benchmarks, we vary $(A_{\cal R}, k_p)$ and use $\Delta = 0.3$ to sets the the width of the enhanced region.  
 
Finally, we note that, if scalar fluctuations are enhanced with a delta function spectral feature at some comoving  wavenumber $k_p$, the previous argument breaks down. For example, if $\mathcal{P}_\mathcal{R}(q) = A_\delta k_p \delta(q - k_p)$, where $A_\delta$ is an arbitrary normalization, then the delta function $\delta(|\vec{k}-\vec{q}| - k_p) = \delta(\cos \theta - k/2k_p)/k$ enforces $\cos\theta = k/2k_p$ and introduces a factor of $1/k$ such that
\bea
    \lim_{k \rightarrow 0} \mathcal{P}_h^{\rm SI}(k) = 4 A_\delta^2 k^2 k_p^2 \, \mathcal{I}_0(k_p)^2 \Theta(2 k_p - k) \propto k^2 \,.
\eea
So a delta function scalar power spectrum yields an unphysical $k^2$ scaling in the IR, a behavior which ultimately traces back to a divergence at $k = k_p$. However, an infinitely sharp peak is not physical, and therefore violates one of the necessary ECT requirements discussed in Appendix~\ref{app:UETC}. For a narrow but finite-width peak, Ref.~\cite{Pi:2020otn} demonstrates that the tensor power spectrum exhibits a $k^2$ scaling in the near-IR before eventually transitioning to the standard $k^3$ behavior in the deep-IR. Any physical ansatz for $\mathcal{P}_\mathcal{R}$ will eventually asymptote to $\mathcal{P}_h(k) \propto k^3$ on sufficiently large scales.

\subsection{Cosmic Strings}\label{subsec:strings}

Phase transitions in the early universe can yield networks of topological defects, such as cosmic strings~\cite{Hindmarsh:1994,Vilenkin:1984ib, Brandenberger:2013npa}, provided the vacuum manifold associated with the symmetry breaking is topologically non-trivial. The resulting string network possesses a highly anisotropic stress-energy tensor, and can therefore source GWs through various string dynamics, including the relativistic oscillations of loops and the formation of cusps and kinks. 

Rather than modeling the full string network, which is highly non-linear and computationally expensive to simulate, it is customary to adopt effective descriptions like the Unconnected Segment Model (USM)~\cite{Albrecht:1997,Pogosian:1999}. In this description, one approximates the network as a collection of uncorrelated, randomly-oriented string segments with length $L$ and RMS velocity $v$. The continuous creation and removal of segments mimics the evolution of a scaling string network, and by calibrating phenomenological parameters to numerical simulations, the USM can accurately and efficiently reproduce the statistical properties of a real network relevant for cosmological observables.

On large scales, the string network has a random walk structure with a characteristic correlation length $L$, which defines the physical length of string segments in the USM. Each segment also has a velocity $v$ and curvature parameter $\kappa$ which, along with $L$, obey equations of motion following from the Nambu-Goto action with a phenomenological interaction term~\cite{Martins:2000}. We also introduce the parameter $\alpha$, which captures the contribution from small-scale structure to string energy and tension. In the scaling regime, there exists a generic attractor solution where each of these quantities assumes a particular functional form -- see Ref.~\cite{Avgoustidis:2012}. In particular, the comoving correlation length $\lambda \equiv L/a$ becomes
$
    \lambda = \xi \tau \,,
$
where $\xi \simeq 0.13$ during radiation domination~\cite{Pogosian:1999}.

\begin{figure}
\includegraphics[width=0.65\textwidth]{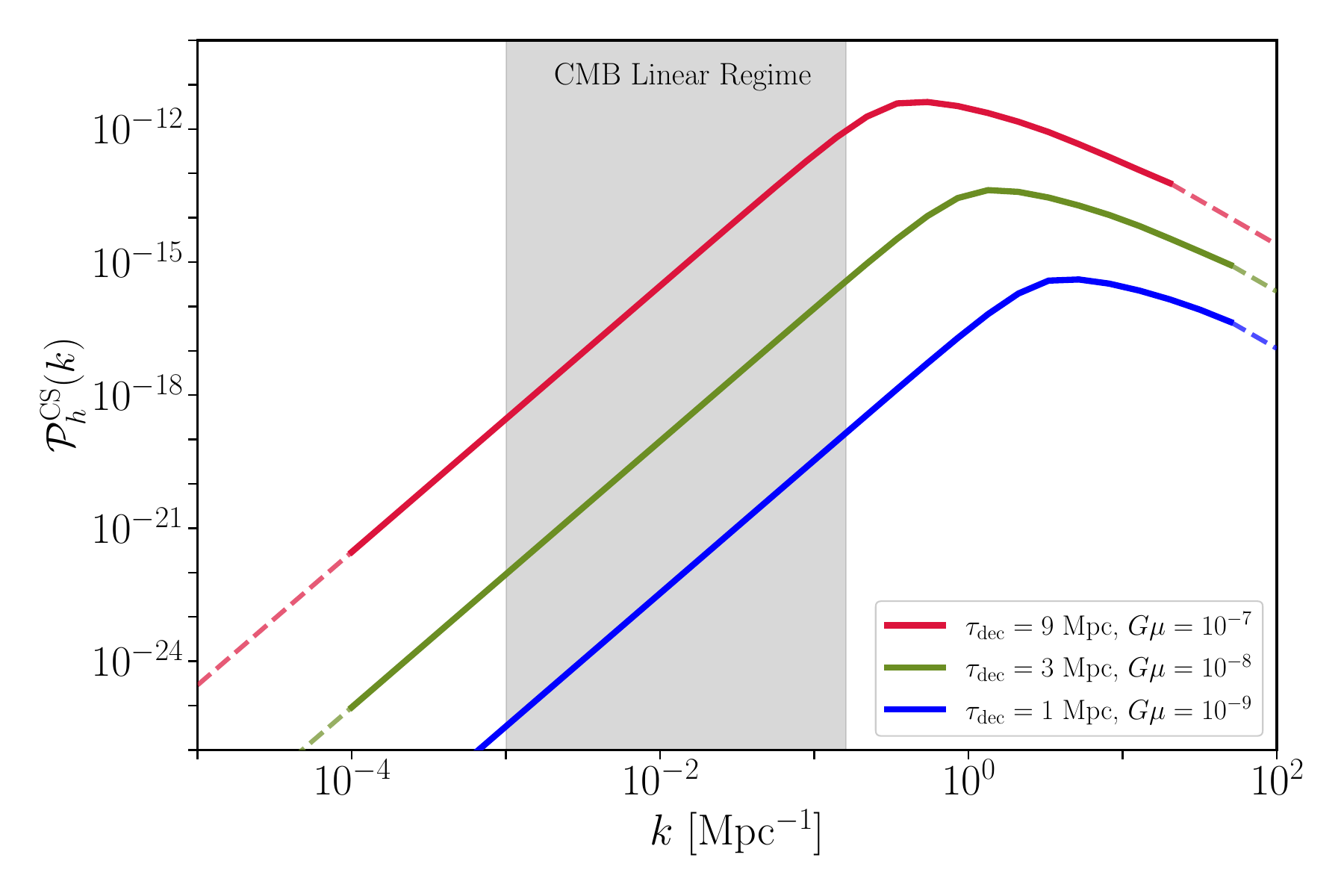}
\caption{Sample dimensionless tensor power spectrum $\mathcal{P}_h(k)$ from the cosmic-string UETC calculation for three choices of the string decay time: $\tau_{\rm decay}$ = 9, 3, and 1 Mpc (red, green, blue) with three different dimensionless string tensions: $G\mu$ = $10^{-7}$, $10^{-8}$, and $10^{-9}$, respectively. The solid curves are the numerically evaluated spectra from integrating the UETC with fixed network parameters ($\xi$ = 0.13, $v$ = 0.65, $\alpha$ = 1.9) 
over the conformal time interval $\tau \in [\tau_i$, $\tau_{\rm decay}$], where $\tau_i$ = 0.01 Mpc. The gray band marks the approximate CMB linear-regime sensitivity window in $k$. Note that the dashed segments show the expected asymptotic power-law behavior: $k^3$ scaling on large scales and $k^{-2}$ scaling on small scales matched to the numerical results at the ends of each curve. 
}
\label{fig:Strings_Ph}
\end{figure}

During the scaling regime, the string network evolves self-similarly with the expansion of the universe, continuously generating GWs at a constant rate. This turns over to the expected causality-limited scaling $\propto k^3$ at some wavenumber $k_{\rm dec} = 1/\tau_{\rm dec}$ corresponding to the time $\tau_{\rm dec}$ by which all segments have decayed. For example, in the axion string case with $N=1$ domain walls, long strings are pulled by the wall tension and annihilate into axions within a Hubble time once the domain-wall energy dominates over the string tension~\cite{Sikivie:2006ni}. String networks which decay sufficiently early are examples of ECTs. 

The UETC corresponding to this source is \cite{Avgoustidis:2012}
\bea\label{eq:stringUETC}
    P_\Pi^{\rm CS} (\tau_1,\tau_2,k) = \frac{\mu^2}{(1-v^2) k^2} f(\tau_1, \tau_2, \xi, \omega_{\rm dec}) \sum_{i=1}^6 A_i \left[ I_i(\tau_-) - I_i(\tau_+)  \right] \,, 
\eea
where $\tau_\pm = (\tau_1 \pm \tau_2)$ and $\mu$ appears in $U = \alpha \mu$, the physical energy density per unit length of the string. The expressions for $A_i$ and $I_i$ can be found in Ref.~\cite{Avgoustidis:2012}. The function $f$ describes the scaling of the power spectrum due to the evolution of the correlation length $\lambda$, and is a function of $\xi$ and the parameter $\omega_{\rm dec}$, which characterizes how instantaneous the string decay is. The instantaneous decay limit corresponds to $\omega_{\rm dec} \rightarrow 1$, for which
\bea
    f_s(\tau_1, \tau_2, \xi, \omega_{\rm dec} \simeq 1) = \frac{1}{\xi^3 \text{max}[\tau_1,\tau_2]^3} \,.
\eea
The tensor power spectrum is given by Eq.~(\ref{eq:PhUETC}) with $\tau_f \rightarrow \tau_{\rm dec}$ and the UETC~\eqref{eq:stringUETC} above,
\bea\label{eq:Phstringinprogress}
    P_h^{\rm CS}(k) \!= \bigg(\! \frac{16 \pi G}{a_{\rm dec}} \!\bigg)^2 \! \int_{\tau_i}^{\tau_{\rm dec}} \!\! d\tau_1 \int_{\tau_i}^{\tau_{\rm dec}} \!\! d\tau_2 \, a(\tau_1)^3 a(\tau_2)^3 G_k(\tau_{\rm dec}, \tau_1) G_k(\tau_{\rm dec}, \tau_2) P_\Pi^{\rm CS}(\tau_1, \tau_2, k) \,.
\eea

 Here we derive analytically the small-wavenumber limit to show the emergence of the causality-limited white-noise scaling. For $k \tau_{\rm dec} \ll 1$, Eq.~(\ref{eq:stringUETC}) becomes to leading order~\cite{Avgoustidis:2012} 
\bea
    \lim_{k \rightarrow 0} P_\Pi^{\rm CS} (\tau_1,\tau_2,k) = \frac{\mu^2 [1 + v^2 (\alpha^2 - 2) + v^4 ( 1 - \alpha^2 + \alpha^4)]}{15 \alpha^2 \xi (1-v^2)} \left( \frac{\tau_1 \tau_2}{\text{max}[\tau_1,\tau_2]^3} \right) 
\eea
which is manifestly $k$-independent. Substituting this expression into Eq.~(\ref{eq:Phstringinprogress}) and using also the small-$k$ expansion of the Green's function from Eq.~\eqref{eq:Gsmallk} yields the dimensionless tensor power spectrum
\bea
    \lim_{k \rightarrow 0} \mathcal{P}_h^{\rm CS}(k) = \frac{k^3}{2\pi^2} \bigg(\! \frac{16 \pi G}{a_{\rm dec}} \!\bigg)^2 \! \int_{\tau_i}^{\tau_{\rm dec}} \!\! d\tau_1 \int_{\tau_i}^{\tau_{\rm dec}} \!\! d\tau_2 \, a(\tau_1)^3 a(\tau_2)^3 (\tau_{\rm dec} - \tau_1) (\tau_{\rm dec} - \tau_2) P_{\Pi}^{\rm CS}(\tau_1, \tau_2, 0) \propto k^3 \,,
\eea
which indeed scales as white noise in the deep infrared. 
In Fig. \ref{fig:Strings_Ph} we evaluate ${\cal P}_h^{\rm CS}$ numerically for some representative benchmarks. As in the previous examples, these parameter choices result in the characteristic $k^3$ scaling on CMB scales, as expected from the analytical expansion above.

\section{Conclusion}

In this paper we have identified a universal scaling property of all post-inflationary sources 
that generate GW from sub-horizon physics before the CMB era. Due to causality, these so-called early causal tensor (ECT) sources all predict white-noise tensor power spectra (${\cal P}_h \propto k^3$) on the large scales observed in the CMB, independently of their microphysical properties. Since the tensor power spectrum governs CMB observables, ECTs also predict the same multipole spectral shape for the $B$-mode angular power spectrum. In analogy with slow-roll inflationary $B$-mode predictions, ECT power spectra differ only in terms of their overall amplitude, which depends on model-specific details. 
Since these sources exhibit $\propto k^3$ dependence in their power spectra, they are not scale invariant and differ from inflationary tensor modes by predicting enhanced power on small scales and suppressed power on large scales. Consequently, the ECT $B$-mode angular signal peaks near $\ell \sim 1000$, whereas the inflationary peak is around $\ell \sim 100$, so these two source classes (ECTs and inflation) are distinguishable with sufficiently precise measurements on different angular scales.

Using this universal scaling behavior, we have introduced a unified description of all ECT sources in terms of a parameter $r_{\rm ect}$ defined in analogy with the inflationary tensor-to-scalar ratio, which quantifies the amplitude of an ECT $B$-mode prediction. This description can also facilitate experimental analyses to probe ECT sources using $B$-mode data and, in a companion paper~\cite{MainPaper}, we present the first limits on these sources. Furthermore, the same parametrization in terms of $r_{\rm ect}$ can  characterize the infrared tail of stochastic GW spectral densities that arise from ECT sources. Since the CMB is sensitive to the lowest measurable GW frequencies ($f \sim 10^{-17}$ Hz), this offers a complementary probe of GW spectral densities beyond the reach of PTA and interferometer probes.  

Finally, we have considered three representative case studies of ECT sources: cosmological phase transitions \cite{Greene:2024},
cosmic string networks \cite{Avgoustidis:2012}, and scalar-induced tensor modes \cite{Ireland:2025}. In each example, we have taken the small $k$ limit and identified the regime in which each source exhibits ECT behavior across the full CMB sensitivity window.

\begin{acknowledgments}
We are grateful to C.L. Reichardt and J.A. Zebrowski for valuable conversations and feedback on an earlier version of this draft. AI is supported by NSF Grant PHY-2310429, Simons Investigator Award No.~824870, DOE HEP QuantISED award \#100495, the Gordon and Betty Moore Foundation Grant GBMF7946, and the U.S.~Department of Energy (DOE), Office of Science, National Quantum Information Science Research Centers, Superconducting Quantum Materials and Systems Center (SQMS) under contract No.~DEAC02-07CH11359. This document was prepared using the resources of
the Fermi National Accelerator Laboratory (Fermilab),
a U.S. Department of Energy, Office of Science, Office
of High Energy Physics HEP User Facility. Fermilab
is managed by Fermi Forward Discovery Group.
YT is supported by the NSF Grant PHY-2412701. YT would also like to thank the host of Fermilab URA Visiting Scholar Program, supported by the URA-22-F-13 fund.
\end{acknowledgments}

\appendix
\section{IR Limit of UETC}\label{app:UETC}

In this appendix, we justify the claim of Eq.~(\ref{eq:UETCk0lim}) that the unequal time correlator (UETC) of a physical causal source becomes $k$-independent in the limit $k \rightarrow 0$. While Ref.~\cite{Cai:2019} rigorously proves this behavior for sources whose stress-energy tensors are bilinear in scalar and vector fields, this ansatz is too restrictive to encompass the broad class of sources considered here. We show that, more generally, any causal source whose stress tensor has finite energy density, is locally conserved, is active for only a finite time, and whose correlations decay sufficiently rapidly on scales greater than the causal horizon satisfies this limiting behavior. 

Concretely, we seek to show that for such a source
\bea\label{eq:UETC:prove}
    \lim_{k \rightarrow 0} P_\Pi(\tau_1, \tau_2, k) \equiv P_\Pi(\tau_1, \tau_2, 0) < \infty \,.
\eea
In general, the UETC can be expressed in terms of the 2-point function of real-space source terms as
\bea\label{eq:UETC:realspace}
    P_\Pi(\tau_1, \tau_2, k) = \int d^3 r \, e^{- i \vec{k} \cdot \vec{r}} \langle \Pi_\lambda(\tau_1, \vec{0}) \Pi_\lambda(\tau_2, \vec{r}) \rangle \,,
\eea
where we have assumed homogeneity and isotropy. Combining Eqs.~(\ref{eq:UETC:prove}) and (\ref{eq:UETC:realspace}) gives the inequality
\bea\label{eq:UETC:condition}
    \int d^3 r \, \langle \Pi_\lambda(\tau_1, \vec{0}) \Pi_\lambda(\tau_2, \vec{r}) \rangle < \infty \,.
\eea
By causality, any correlations between $\Pi_\lambda(\tau_1,\vec{0})$ and $\Pi_\lambda(\tau_2,\vec{r})$ must originate from events lying in the intersection of their past light cones. This intersection has a characteristic size which is at most that of the smaller horizon,\footnote{This is an upper bound. The actual intersection decreases with increasing $r$.} $\ell_{\rm corr} \lesssim \text{min}(\tau_1, \tau_2)$. For a source which is only active for a finite duration, $\tau_1$ and $\tau_2$ are bounded and the correlation length remains finite. Correlations may be non-zero for $r \lesssim \ell_{\rm corr}$, but must decay for $r \gg \ell_{\rm corr}$. This motivates decomposing the integral~(\ref{eq:UETC:condition}) into contributions from $r < \ell_{\rm corr}$ and $r > \ell_{\rm corr}$. The contribution coming from the region $r < \ell_{\rm corr}$ is automatically finite for any physical source with bounded energy density. To ensure that the contribution from $r > \ell_{\rm corr}$ converges,\footnote{Although cluster decomposition in combination with the isotropy and homogeneity of the background ensures that $\lim_{r \rightarrow \infty} \langle \Pi_\lambda(\tau_1, \vec{0}) \Pi_\lambda(\tau_2, \vec{r}) \rangle = 0$, a stronger condition is needed here.} it is sufficient that the UETC decay faster than $1/r^3$ on scales larger than the correlation length. This is a mild assumption which is satisfied by all the ECT case studies considered here.

\bibliography{apssamp}

\end{document}